\documentclass[aps,pra,twocolumn,superscriptaddress]{revtex4-1}
\usepackage{graphicx}
\usepackage{amssymb}
\usepackage{amsmath}
\usepackage{bm}
\usepackage{xcolor}
\usepackage{epstopdf}
 \usepackage{verbatim}

\begin{document}

\title{Large-area $^{87}$Rb Bose-Einstein condensate in a clipped-Gaussian optical dipole trap}
\author{Younghoon Lim}
\affiliation{Department of Physics and Astronomy, Seoul National University, Seoul 08826, Korea}

\affiliation{Center for Correlated Electron Systems, Institute for Basic Science, Seoul 08826, Korea}

\author{Junhong Goo}
\affiliation{Department of Physics and Astronomy, Seoul National University, Seoul 08826, Korea}

\author{Haneul Kwak}
\affiliation{Department of Physics and Astronomy, Seoul National University, Seoul 08826, Korea}

\author{Y. Shin}
\email{yishin@snu.ac.kr}

\affiliation{Department of Physics and Astronomy, Seoul National University, Seoul 08826, Korea}

\affiliation{Center for Correlated Electron Systems, Institute for Basic Science, Seoul 08826, Korea}

\affiliation{Institute of Applied Physics, Seoul National University, Seoul 08826, Korea}

\begin{abstract}
We demonstrate a production of large-area $^{87}$Rb Bose-Einstein condensates (BECs) using a non-Gaussian optical dipole trap (ODT). The ODT is formed by focusing a symmetrically truncated Gaussian laser beam and it is shown that the beam clipping causes the trap geometry elongated and flattened along the beam axis direction. In the clipped-Gaussian ODT, an elongated, highly oblate BEC of $^{87}$Rb is generated with length and width of approximately 470~$\mu$m and 130~$\mu$m, respectively, where the condensate healing length is estimated to be $\xi\approx 0.25~\mu\textrm{m}$ at the trap center. The ODT is characterized to have a quartic trapping potential along the beam axis and the atom density of the condensate is uniform within 10\% over $1000\xi$ in the central region. Finally, we discuss the prospect of conducting vortex shedding experiments using the elongated condensate.
\end{abstract}


\maketitle
 
\section{Introduction}

An optical dipole trap (ODT) is a popular trapping method for cold neutral atoms~\cite{Grimm}, and is typically created by focusing a far-off-resonant Gaussian laser beam. In contrast to a magnetic trap, an ODT can confine atoms regardless of their spin states, thus allowing the study of spin dynamics with the trapped samples~\cite{Stamper}. In addition, its trapping geometry can be tailored to some extent, e.g., by engineering the laser beam profile~\cite{Wright,Gauthier_dmd}, using multiple laser beams~\cite{Gaunt,Hueck} and particularly, their spatial interference to provide lattice potentials~\cite{Bloch}, or rapidly scanning a laser beam to generate a time-averaged potential~\cite{Bell}. By virtue of these merits, diverse ODTs with many different geometries have been designed and utilized over the last decades in the cold atom experiments to vastly expand their research scope.

In this paper, we present a simple variation of single-beam ODT that enables the production of an atomic sample with large area. The experimental setup for the ODT, where a collimated, elliptical laser beam is symmetrically truncated by a horizontal slit and focused through a cylindrical lens, is illustrated in Fig.~1. When a truncated Gaussian laser beam is focused, the focal region is elongated and furthermore, the beam waist becomes uniform over the long focal region~\cite{Gillen}. The flattened laser beam has been successfully used in bio-imaging as an optical sheet to selectively excite a slice region of a sample~\cite{Olarte18}. In this work, we employ a non-Gaussian laser beam as the ODT and demonstrate the generation of large-area, highly oblate Bose-Einstein condensates (BEC) of $^{87}$Rb. In an optimal clipping condition, we obtain a condensate whose length and width are approximately 470~$\mu$m and 130~$\mu$m, respectively. The condensate thickness is $\approx 11\xi$, where $\xi\approx 0.25~\mu\textrm{m}$ is the condensate healing length at the trap center. We observe that the ODT confinement along the beam axis is well described by a quartic potential and that the atom density is uniform within 10\% over the half of the condensate in the central region.

\begin{figure}[t]
 \includegraphics[width=80mm]{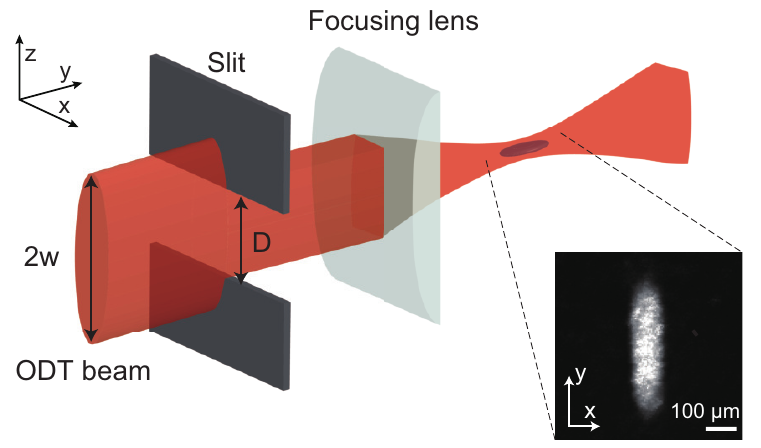}
  \caption{Clipped-Gaussian optical dipole trap (ODT). An elliptical laser beam is symmetrically truncated by a horizontal slit and focused through a cylindrical lens to form an ODT. $2w$ and $D$ denote the initial beam diameter and the opening width of the slit, respectively. The inset is an image of a $^{87}$Rb Bose-Einstein condensate (BEC) trapped in the ODT.}
\label{fig1}
\end{figure}

The large-area, highly oblate BEC is expected to be beneficial to study the critical vortex shedding dynamics~\cite{Frisch,Winiecki,Neely_vd,Kwon_crit} and related turbulence phenomena in a superfluid~\cite{Neely_qt,Kwon_qr,Seo,Gauthier_gv,Johnstone_gv}. In classical fluid dynamics, the Strouhal number is defined as $\textrm{St} = f_v D/v$, where $f_v$ is the vortex shedding frequency, $D$ is the characteristic diameter of the obstacle, and $v$ is its moving speed, and it is well known that $\textrm{St}\approx 0.2$ over a large range of the Reynolds number~\cite{Leinhard}. Intriguingly, similar universal behavior of $\textrm{St}$ was predicted for a superfluid in a numerical study~\cite{Reeves} and recently, a tentative experimental evidence was reported~\cite{Kwon_vs}. An elongated and flattened BEC provides a long moving distance for an obstacle, allowing better measurements of the vortex shedding frequency. We emphasize that the beam clipping method can be easily implemented in experiments, providing a simple and practical way to elongate a trapped sample.

The remainder of this paper is organized as follows. In Sec.~II, we investigate the ODT elongation and flattening effect due to the laser beam clipping by directly measuring the beam intensity distribution of the focused laser beam and by performing numerical simulations. In Sec.~III, we present our experimental results of generating a large-area Bose-Einstein condensate and the characterization of the trapping potential. Finally, a summary is provided in Sec.~IV, together with outlook on the vortex shedding experiment with the large-area sample.

\section{Clipped-Gaussian optical dipole trap}

We first investigate the effect of laser beam clipping using a separate optics setup that emulates the ODT in our BEC experiment. In the setup, we use a 780-nm, elliptical Gaussian laser beam, whose horizontal and vertical $1/e^2$ diameters are 2 and 4~mm, respectively. As depicted in Fig.~1, the laser beam is symmetrically truncated by a horizontal slit and focused through a cylindrical lens with a focal length $f=100$~mm. We measure the intensity profile of the focused laser beam as a function of the axial position $y$ near the focal spot, from which mapping the three-dimensional (3D) intensity distribution, $I(x,y,z)$, that is proportional to the trapping potential for atoms. The intensity profile along the $x$ direction was observed to maintain its original Gaussian form, and in the following, we consider only the two-dimensional (2D) intensity distribution $I(y,z)$.

Fig.~2(a) displays the intensity distributions $I(y,z)$ measured for various clipping conditions. Here the intensity is normalized by its peak value $I_0$ at the focus and the clipping condition is parameterized with $\gamma=D/2w$, the ratio of the slit width $D$ to the $1/e^{2}$ diameter $2w$ of the incident laser beam. It is clearly shown that the focal region is elongated along the beam propagation direction as $\gamma$ decreases, i.e., the laser beam becomes more clipped. For $\gamma = 0.5$, the high-intensity region, where $I/I_0>0.9$, is stretched over 6~mm along the $y$-axis, which is approximately 4 times longer than that without clipping. The focus position is slightly shifted towards the focusing lens with decreasing $\gamma$~\cite{Gillen}, which is attributed to the spherical aberration of the focusing lens.

In Fig.~2(b), we plot the $1/e^2$ radius of the focused beam as a function of the axial position $y$. When the laser beam is significantly truncated for $\gamma < 1$, the beam radius noticeably increases and furthermore, it exhibits peculiar $y$-dependence such that there appears a central region with a quasi-constant beam radius. This means that the resultant ODT would be not only elongated but also flattened at its center by the laser beam clipping.

\begin{figure}[t]
    \centering
    \includegraphics[width=84mm]{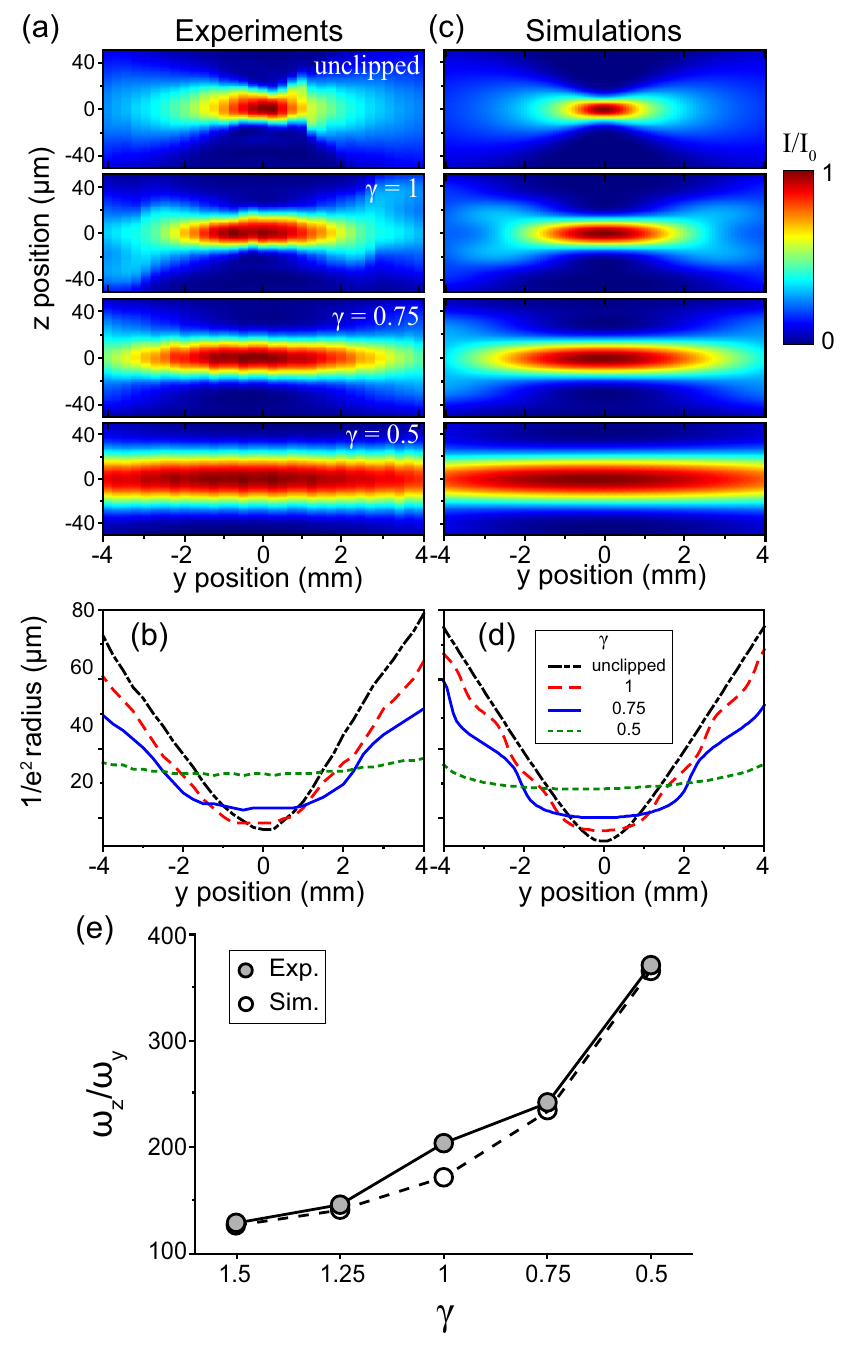}
    \caption{Evolution of an ODT for the laser beam clipping. (a) Intensity distribution $I(y,z)$ of the focused laser beam for various clipping conditions, measured in a separate optics setup (see text for details). $I_0$ denotes the peak value at the focus and $\gamma=D/2w$ is the relative opening width of the slit. (b) $1/e^2$ radius of the focused laser beam as a function of the axial position $y$. (c) Numerical simulation results of $I(y,z)$ and (d) the $1/e^2$ radius for the experimental conditions in (a) and (b). (e) Trapping frequency ratio, $\omega_z/\omega_y$, as a function of $\gamma$. $\omega_{y(z)}$ is the trapping frequency of the ODT in the $y(z)$ direction at its center region of $I/I_0>0.8$.}
    \label{fig:experiment}
\end{figure}

We also investigate the clipped-Gaussian ODT by numerically calculating the propagation of the clipped and focused laser beam. In scalar diffraction theory, the electric field of the laser beam, $U_f(z,d)$, after propagating by a distance $d$ from the lens is given by
\begin{equation}
    U_f(z,d) = \dfrac{1}{\sqrt{\lambda d}}\int{t(z')\theta_s(z')U_i(z') e^{ -\frac{i \pi}{\lambda d}  (z-z')^2 } dz'}
    \label{green}
\end{equation}
in the form of the Rayleigh-Sommerfeld diffraction integral using a paraxial approximation, where $U_i(z) = U_0 \exp(-\frac{z^2}{w^2})$ is the electric field of the incident Gaussian beam, $\theta_s(z)=\theta(z+\frac{D}{2})\theta(\frac{D}{2}-z)$ with $\theta(z)$ being the Heaviside step function represents the truncation by the slit, and $t(z)=\exp( \frac{i \pi }{\lambda f} z^2)$ is the transmission function of the lens for wavelength $\lambda$. Here we neglect the propagation from the slit to the lens. Based on this equation, we numerically calculate the intensity distribution $I_\textrm{sim}(y,z)=|U_f(z,d=f+y)|^2$ for our experimental condition [Fig.~2(c)] and determine the evolution of the $1/e^2$ radius of the focused laser beam [Fig.~2(d)]. We find the numerical results in good quantitative agreement with the experimental data.

The elongation due to the laser beam clipping is further characterized by measuring the trapping frequency ratio, $\alpha=\omega_z/\omega_y$, of the clipped-Gaussian ODT as a function of $\gamma$, where $\omega_{y(z)}$ is the trapping frequency of the ODT in the $y(z)$ direction at its center. The value of $\alpha$ is determined from a fit of an inverted 2D quadratic function of $I_\textrm{fit}(y,z)= I_0[1- \frac{1}{2}\kappa^2 ( y^2 + \alpha^2 z^2)]$ to the measured $I(y,z)$ for the high-intensity region of $I/I_0>0.8$ and the results are displayed in Fig.~2(e). For $\gamma=0.5$, $\alpha$ increases up to $\approx 400$. To put this value in perspective, we compare it to the trapping frequency ratio of a normal Gaussian ODT, which is given by $\alpha = \frac{\sqrt{2}\pi w_0}{\lambda}= \frac{\sqrt{2}f}{w}$, where $w_0= \frac{\lambda}{\pi} \frac{f}{w}$ is the $1/e^2$ beam radius at the focus. In order to obtain such a high value of $\alpha\approx 400$ in our optics setup without clipping, the beam radius $w_0$ needs to be larger than $70~\mu$m, which can be achieved by, e.g., reducing the incident beam diameter to $2w<0.7$~mm. We note that it is practically much easier to clip the laser beam than to reduce the beam diameter, particularly, without affecting the focal position. In the limit of $\gamma \ll 1$, the clipped input beam can be regarded as a flat-top square beam with width $D$ and we find $\alpha \sim \frac{8f}{D}$ from numerical simulations.

\section{Generation of large-area BECs}

\subsection{BEC production}

A schematic view of our apparatus for generating $^{87}$Rb BEC is presented in Fig.~3. The apparatus consists of two vacuum parts: a glass cell to generate a cold atomic beam and a main ultra-high vacuum chamber to produce the BEC. In the glass cell, we form a 2D magneto-optical trap (MOT) and generate an atomic beam by pushing atoms using a red-detuned laser beam to the main chamber. Atoms move through a differential pumping tube, which is 10~cm long with 4~mm inner diameter, and are loaded into a 3D MOT in the main chamber. After full loading, we transfer the atoms into a magnetic quadrupole trap, which is aided by MOT compression and molasses cooling. We then apply rf-induced evaporative cooling to the trapped sample and transfer it into an ODT, avoiding the atom loss due to the Majorana spin flip in the magnetic trap at low temperature~\cite{Heo11}. At the transfer, the number of atoms in the magnetic trap is $\approx 4.2\times10^{8}$, and their temperature is $\approx 17~\mu$K. The sample is further cooled by lowering the trap depth of the ODT and we obtain a quasi-pure BEC containing $\approx 2.2\times 10^{6}$ atoms in the $|F=1, m_{F}= -1\rangle$ state (Fig.~3 inset). The ODT is created by focusing a 1064-nm, elliptical Gaussian beam with $2w= 22$~mm (Fig.~1) and the focal length of the cylindrical lens is 100~mm. The trapping frequencies of the final ODT are $\omega_{x,y,z} \approx 2\pi \times (6.9, 5.1, 225)$~Hz and the condensate is highly oblate with Thomas-Fermi (TF) radii of $R_{x,y,z} \approx (65, 87, 2.0)~\mu$m.

\begin{figure}[t]
 \includegraphics[width=84mm]{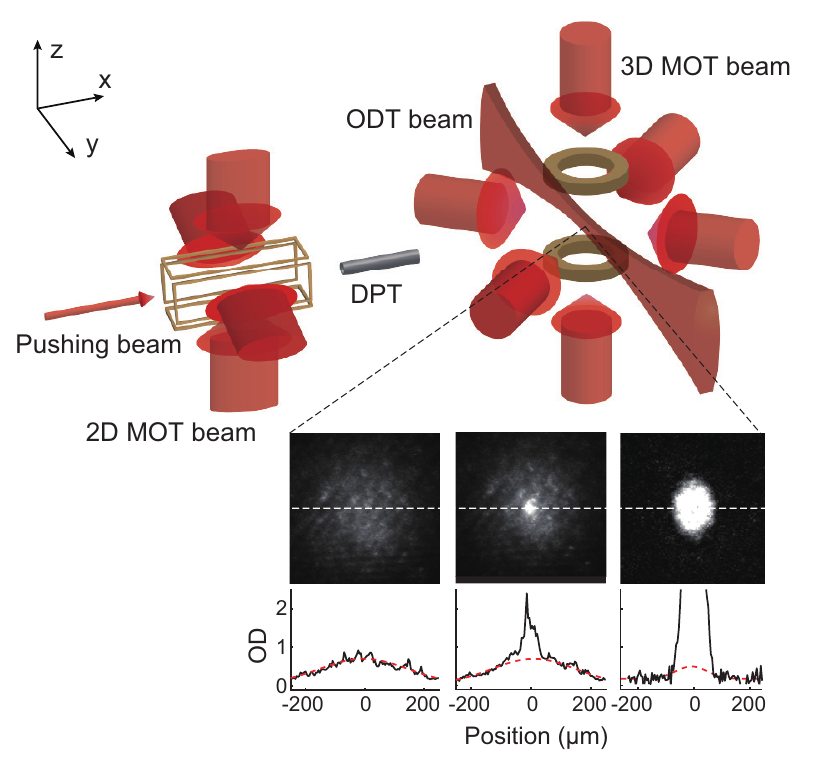}
  \caption{Schematic view of the experimental apparatus for producing a $^{87}$Rb BEC. An atomic beam is generated from a 2D magneto-optical trap (MOT) with a pushing laser beam in a glass cell (not shown). Atoms are loaded into a 3D MOT in a ultra-high vacuum chamber, which are connected to the glass cell via a differential pumping tube (DPT), and subsequently cooled down via evaporation after transferred into a magnetic trap and then in a ODT. The images in the bottom were acquired after a time-of-flight, showing the Bose-Einstein condensation in the atomic cloud. The optical density (OD) profiles along the central dashed lines are displayed below, where the red dashd lines are Guassian curves fit to the outer thermal wings.}
\label{fig4}
\end{figure}

\subsection{Elongation by clipping}

Installing an adjustable horizontal slit before the final focusing lens, we investigate the beam clipping effect by measuring the atom density distribution of the condensate in the ODT for various $\gamma$. Because of the power loss caused by the clipping as well as the volume change of the ODT, it was necessary to adjust the power control of the ODT laser beam during evaporation for each $\gamma$ to maximize the BEC sample. In the tuning process, it was observed that the condensate tends to drift along the beam axis, which is because the axial confinement is weakened by the laser beam clipping and the beam alignment is not perfectly orthogonal to the gravity direction, and that the drifting also depends on the slit position with respect to the beam center. When the condensate is axially moved, it exhibits an unbalanced density profile, indicating the anharmonicity of the axial trapping potential. To compensate the axial drift, we apply a magnetic field gradient along the $y$ direction to balance the axial density distribution of the condensate. The final power of the ODT laser beam is set to maintain the trapping frequency $\omega_{x}$ constant within 10$\%$.

\begin{figure}[t]
 \includegraphics[width=80mm]{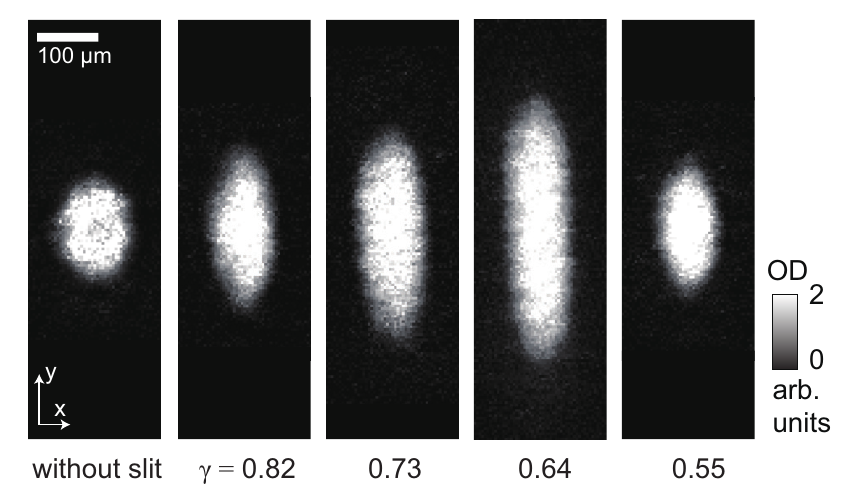}
  \caption{Elongation of the BEC in the clipped-Gaussian ODT. {\it In situ} images of BECs for various $\gamma$ are displayed. The axial extent of the condensate increases along the ODT beam axis with decreasing $\gamma$. The aspect ratio $R_{y}/R_{x}$ is increased to approximately 4 for $\gamma$ = 0.64. Trapping frequency along the $x$ direction, $\omega_{x}$, was maintained within 10\% for all $\gamma$.}
\end{figure}

In Fig.~4, we display the {\it in-situ} images of the trapped condensates for various $\gamma$. It is clearly observed that the condensate is elongated along the beam axis, as expected from the result in the previous section. The aspect ratio $R_y/R_x$ of the sample in the $xy$ plane is measured by determining the TF radius $R_{x(y)}$ in the $x(y)$ direction from an intercept of a linear fit to the outer regions of the {\it in-situ} density profile. The aspect ratio is increased almost to 4 for $\gamma$ = 0.64 with the radius values of $R_{x0,y0}=(63(1), 232(5))~\mu\textrm{m}$, which is a nearly three-fold increase compared to that for the unclipped ODT. We note that the atom number of the condensate is also increased by a factor of five to $\approx 1.16\times 10^{7}$. When the ODT laser beam is further clipped for lower $\gamma < 0.64$, on the other hand, the axial extent of the condensate is markedly decreased together with its atom number reducing. It might be attributed to the low transfer efficiency of atoms to the ODT due to insufficient power of the clipped laser beam or some uncontrolled diffraction effects of the deep clipping, which are not clearly understood at the moment. The power loss is 16$\%$ at $\gamma= 0.64$ and is increased to 25$\%$ for $\gamma= 0.55$.

\subsection{Trap characterization}

\begin{figure}[t]
\includegraphics[width=80mm]{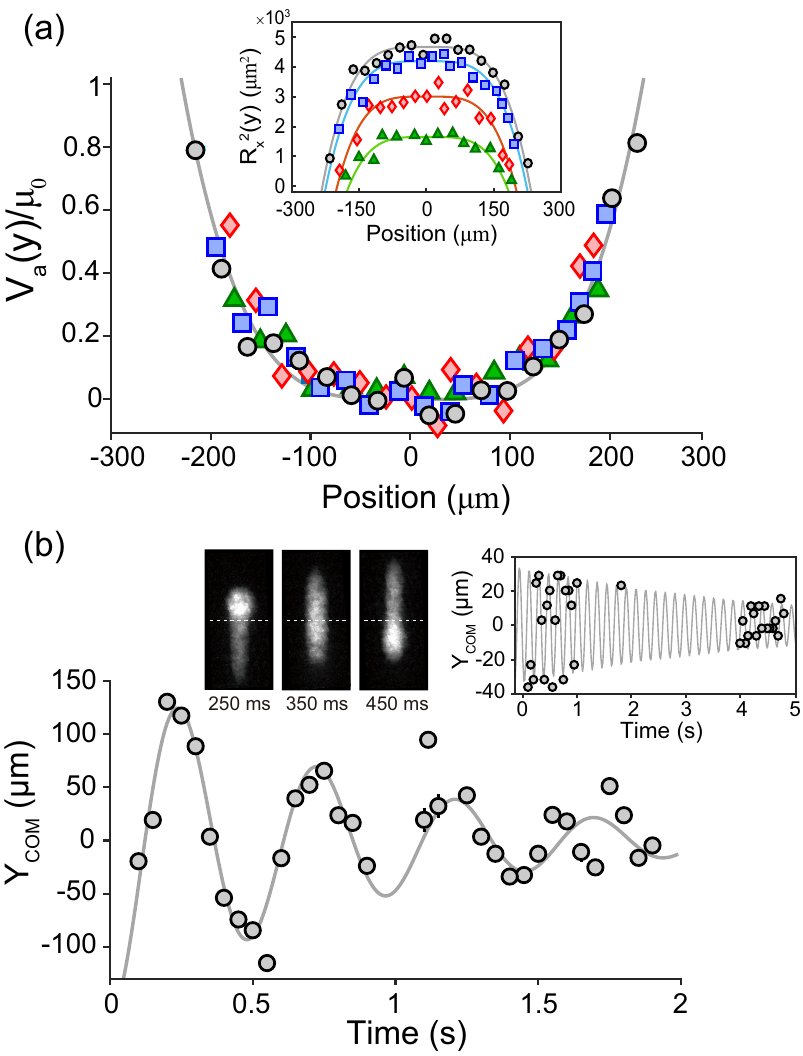}
\caption{Anharmonic trapping potential. (a) The axial trapping potential $V_a(y)$ of the ODT for $\gamma=0.64$ was determined from Eq.~(3) by measuring the Thomas-Fermi radius $R_x(y)$ of the condensate at the axial position $y$. The inset shows the measurement results of $R_x^2(y)$, where different markers denote samples with different atom numbers. $\mu_0$ is the chemical potential of the condensate for the maximum atom number. The gray line indicates a power-law function of $V_{a0}(y)=A |y|^\beta$ with $\beta=3.9$, fit to all the data of $V_a(y)$, and the solid lines in the inset are the corresponding curves for the $R_x^2$ data. Each data point is the mean of fifteen measurements of a same experiment. (b) Damped dipole oscillations. The axial center-of-mass position, $Y_\textrm{COM}$, of the oscillating condensate as a function of time. Each point is the mean of three measurements and it error bar represents their standard deviation. The gray solid line is a damped sinusoidal curve fit to the data. In the upper left, {\it in situ} images of the oscillating condensate are displayed for different times. The right inset shows the oscillation data for the unclipped case.}
\label{fig4}
\end{figure}

We characterize the trapping potential of the clipped-Gaussian ODT at our optimal clipping with $\gamma = 0.64$. The trapping potential is modelled as
\begin{equation}
    V(x,y,z) = \frac{1}{2} m\omega_{x}^{2}x^{2}+V_a(y) + \frac{1}{2} m\omega_{z}(y)^{2}z^{2}
\end{equation}
with $m$ being the atomic mass. In this model, the ODT modifications due to the laser beam clipping are described with a separable potential $V_a(y)$ along the beam axis and the $y$-dependence of $\omega_z$. Here $\omega_x$ is assumed to be independent of the axial position $y$, which is supported by our observation that the condensate undergoes long-lived dipole oscillations in the $x$ direction without significantly distorting its boundary shape, giving $\omega_x=2\pi\times 7.4(1)$~Hz. 

The axial trapping potential $V_a(y)$ is determined from the relation of the chemical potential $\mu$ to the Thomas-Fermi boundary position of the condensate, $\mu=V(R_x(y), y, 0)=\frac{1}{2}m\omega_x^2 R_x^2(0)$, which gives
\begin{equation}
    V_a(y) = \frac{1}{2}m\omega_{x}^{2} \big[ R_x^2(0)-R_{x}^{2}(y) \big],
\end{equation}
where $R_x(y)$ is the Thomas-Fermi radius in the $x$ direction at the axial position $y$. In Fig.~5(a), the axial trapping potential $V_a(y)$ constructed using Eq.~(3) is displayed, where $R_x(y)$ is measured from  {\it in-situ} images of the trapped condensates with various atom numbers [Fig.~5(a) inset]. Fitting a power-law function of $V_{a0}(y)=A |y-y_0|^\beta$ to the measurement results, we find that the trapping potential is quantitatively well described with $\beta = 3.9 \pm 0.1$.

The anharmonicity of the axial trapping potential is demonstrated using the dipole oscillations of the condensate. The oscillations are induced by adiabatically turning on an additional magnetic field gradient to move the condensate by 0.5$R_{y0}$ from the trap center, and suddenly switching it off. The condensate shape changes during the oscillations; the atoms agglomerate periodically, as shown in the images in Fig.~5(b). The oscillations decay rapidly, contrary to the long-lived oscillations in the unclipped ODT [Fig.~5(b) inset].

\begin{figure}[t]
 \includegraphics[width=84mm]{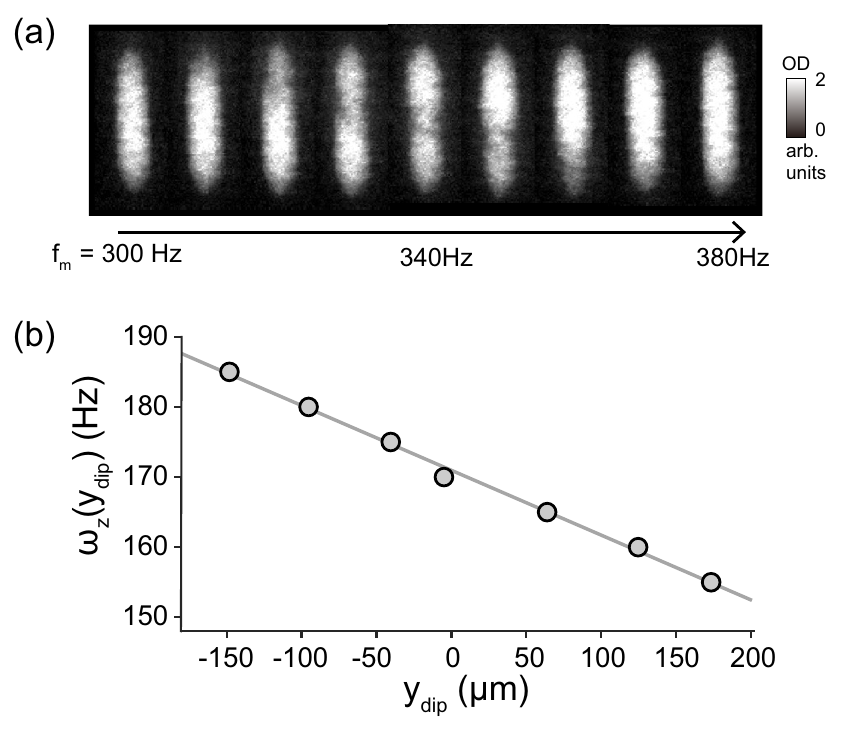}
  \caption{Axial position dependence of the trapping frequency $\omega_z$. (a) {\it In situ} images of BECs after ODT power modulations for 100~ms with variable frequency $f_{m}$. Local density depletion occurs in the condensate due to parametric heating for $f_{m}=2\omega_z$ at the density dip position. (b) The density dip position $y_\textrm{dip}$ was determined from a Gaussian fit to the axial density profile obtained by integrating the image along the $x$ axis. The solid line indicates a linear fit to the data.}
\label{fig4}
\end{figure}

The trapping frequency $\omega_{z}$ along the tight confining direction is measured via parametric heating by sinusoidally modulating the ODT beam power. The modulations are applied for 100~ms and their amplitude is 5\% of the final ODT beam power. As we scan the modulation frequency $f_{m}$, a density dip appears in the sample and its axial position changes with varying $f_{m}$ [Fig.~6(a)]. The density dip formation results from local heating by the trap modulations for the resonance of $f_{m}=2\omega_z$ at the dip position. We find that $\omega_z$ exhibits a small linear dependence on the axial position [Fig. 6(b)], and a linear function fit to the measurement results gives $\omega_{z}(y)=\omega_{z0}(1+By)$ with $\omega_{z0}= 2\pi\times 171(1)$~Hz, $B=5.4(3) \times 10^{-4}~\mu\mathrm{m}^{-1}$.

Putting together all the trap characterization results, we describe the trapping potential $V(x,y,z)$ as
\begin{equation}
    V(x,y,z) = \frac{1}{2}m[\omega_{x}^{2}x^{2}+\omega_{z0}^{2}(1+By)^{2}z^{2}]+ A\vert y \vert^{\beta}
\end{equation}
with $\beta=3.9$ and $A=\mu_0/R_{y0}^{\beta}$, where $\mu_0=\frac{1}{2}m\omega_x^2 R_{x0}^2$ is the chemical potential of the condensate for the maximum atom number in our experiment. For $\mu_0=k_{B}\times~45.7$~nK, the atom number $N_0$ of the condensate is calculated by numerically integrating the atom density $n(\textbf{r}) = [\mu_0-V(\textbf{r})]/U_{0}$ over the sample region, where $U_{0} = \frac{4\pi\hbar^{2}a_s}{m}$ with $a_s$ being the $s$-wave scattering length, yielding $N_0 \approx 1.1\times 10^{7}$, consistent with the measured value. The condensate healing length is estimated to be $\xi = \hbar/\sqrt{2m\mu_0} \approx 0.25 ~\mu\textrm{m}$ at the trap center, and in units of $\xi$, the length and width of the condensate are expressed as $2R_{y0}/\xi \approx 1880$ and $2R_{x0}/\xi \approx 513$, respectively.

\section{Summary and outlook}

We described the clipped-Gaussian ODT which is formed by focusing a symmetrically truncated Gaussian beam, and presented it as a simple and practical method for enlarging the spatial extent of a trapped atomic sample. We generated large-area Bose-Einstein condensates in the ODT and showed that the trapping potential along the beam axis direction is well described to be quartic so that the atom density is uniform within 10\% over the half of the condensate along the elongated direction.

The large-area BEC would be beneficial to many experiments, among which our immediate interest is investigating the vortex shedding dynamics. Vortex shedding or wake generation behind a moving obstacle has been widely studied in classical fluids, and its extension to a superfluid was recently performed with atomic BEC systems using a focused laser beam as an optical obstacle~\cite{Neely_vd,Kwon_crit,Kwon_vs,Kwon_ps}. In experiments with penetrable obstacles~\cite{Kwon_ps}, periodic shedding of vortex dipoles was observed and the linear relationship between the shedding frequency $f_v$ and the obstacle velocity $v$ was demonstrated as $f_{v}=a (v-v_{c})$ with $v_{c}$ being the critical velocity for vortex shedding~\cite{Frisch,Winiecki}. For impenetrable obstacles, the observation of von Kármán vortex streets was reported~\cite{Kwon_vs}. Moreover, it seemed that the number of vortex clusters, $N_c$, shed for a fixed travel distance $L$ tends to be saturated with increasing the obstacle velocity, which intriguingly suggests a constant Strouhal number as $\textrm{St}=f_v D/v\approx (N_c/2L) D$ like observed in classical fluids. Assuming constant $\textrm{St}$, it might be speculated that the proportionality constant $a$ of $f_v$ for a penetrable obstacle is understood as $a=\textrm{St}/D_\textrm{eff}$ with $D_\textrm{eff}$ being the effective diameter of the obstacle. The elongated BEC prepared in the clipped-Gaussian ODT provides an improved setting for the vortex shedding experiments, allowing a longer travel distance with smaller atom density variations. In the previous experiment of Ref.~\cite{Kwon_vs}, the maximum stirring distance was about $300\xi$, resulting in $N_{c}\leq 4$, preventing precise determination of the shedding frequency. In our sample, the travel distance can be stretched to $L> 1000\xi$ and a linear extrapolation predicts $N_c> 10$ for high $v$. We expect that the improvement will facilitate a quantitative study of the vortex shedding dynamics in a BEC, providing an interesting opportunity to explore its possible universality which would establish the superfluid Reynolds number~\cite{Barenghi, Finne, Reeves}.

\begin{acknowledgments}
This work was supported by the National Research Foundation of Korea (NRF-2018R1A2B3003373, NRF-2019M3E4A1080400) and the Institute for Basic Science in Korea (IBS-R009-D1).
\end{acknowledgments}

\end{document}